\documentstyle{article}
\begin{document}

\begin{center}
{\LARGE \bf Orbital variations and outbursts of the 
\vspace{1ex}

unusual variable star
V1129~Centauri}\footnote{Based partially on observations taken at the 
Observat\'orio do Pico dos Dias / LNA}

\vspace{1cm}

{\Large \bf Albert Bruch}

\vspace{0.5cm}
Laborat\'orio Nacional de Astrof\'{i}sica, Rua Estados Unidos, 154, \\
CEP 37504-364, Itajub\'a - MG, Brazil
\vspace{1cm}

(Published in: New Astronomy, Vol.\ 57, p.\ 51 -- 58 (2017))
\vspace{1cm}
\end{center}

\begin{abstract}
The variable star V1129~Cen is classified in the GCVS as being of 
$\beta$~Lyr type. Unusual for such stars, it exhibits outbursts 
roughly once a year, lasting for $\sim$40 days. For this reason, a
relationship to the dwarf novae has been suspected. Here, for the
first time a detailed analysis of the light curve of the system is
presented. Based on observations with high time resolution obtained
at the Observat\'orio do Pico dos Dias and on the long term ASAS light
curve the orbital variations of the system are studied. They are dominated
by ellipsoidal variations and partial eclipses of a probably slightly 
evolved F2 star in a binary with an orbital period of 
$21^{\raisebox{.3ex}{\scriptsize h}}$ $26^{\raisebox{.3ex}{\scriptsize m}}$.
Comparison with the characteristics of dwarf novae show 
that the observational properties of V1129~Cen can be explained if it is
just another dwarf novae, albeit with an unusually bright and early type
mass donor which outshines the accretion disk and the mass gainer
to a degree that many normal photometric and spectroscopic hallmarks of
cataclysmic variables remain undetected.

\phantom{.}

{\parindent0em Keywords:
Stars: binaries: eclipsing -- 
Stars: variables: general -- 
Stars: novae, cataclysmic variables -- 
Stars: dwarf novae --
Stars: individual: V1129~Cen}
\end{abstract}

\section{Introduction}
\label{Introduction}

Cataclysmic variables (CVs) are binary stars where a Roche-lobe filling 
late-type component (the secondary) transfers matter via an accretion disk 
to a white dwarf primary. A particular subclass of CVs are the dwarf novae
which occasionally exhibit outbursts with amplitudes of a few magnitudes,
lasting from days to weeks. These are caused by a temporary increase of the
brightness of the accretion disks in these systems.

It may be surprising that even after decades of 
intense studies of CVs there are still an appreciable number of known or 
suspected systems, bright enough to be easily observed with comparatively small
telescopes, which have not been studied sufficiently for basic parameters
to be known with certainty. In some cases even 
their very class membership still requires confirmation. 

Therefore, I started a small observing project aimed at a better understanding
of these stars. First results have been published by 
Bruch (2016, 2017a) and Bruch \& Diaz (2017). Here, I present time 
resolved photometry and a limited amount of spectroscopy of the unusual
system V1129~Cen. To these data I add long term observations 
retrieved from the ASAS-3 data archive (Pojmanski 2002).

V1129~Cen is not a normal CV. In fact, 
the relationship of the star to the cataclysmic variables is quite unclear.
In spite of its high brightness of 
$\sim$$9^{\raisebox{.3ex}{\scriptsize m}}_{\raisebox{.6ex}{\hspace{.17em}.}}7$
not many details
are known about the star. It is classified as a $\beta$~Lyr 
type eclipsing binary in the $17^{\rm th}$ name list of variable stars 
(Kazarovetz et al.\ 2008). 
$\beta$~Lyr systems are binaries made up of stars in tight
or even semi-detached orbits. Their evolutionary state may range from two main 
sequence stars to a pair with a highly evolved secondary component and a
less evolved primary with mass transfer between them (Hoffman et al.\ 2008). 
Due to the proximity of the stellar components the light curves are 
dominated by ellipsoidal variations often in combination with mutual 
eclipses. 

\begin{figure}
   \parbox[]{0.1cm}{\epsfxsize=14cm\epsfbox{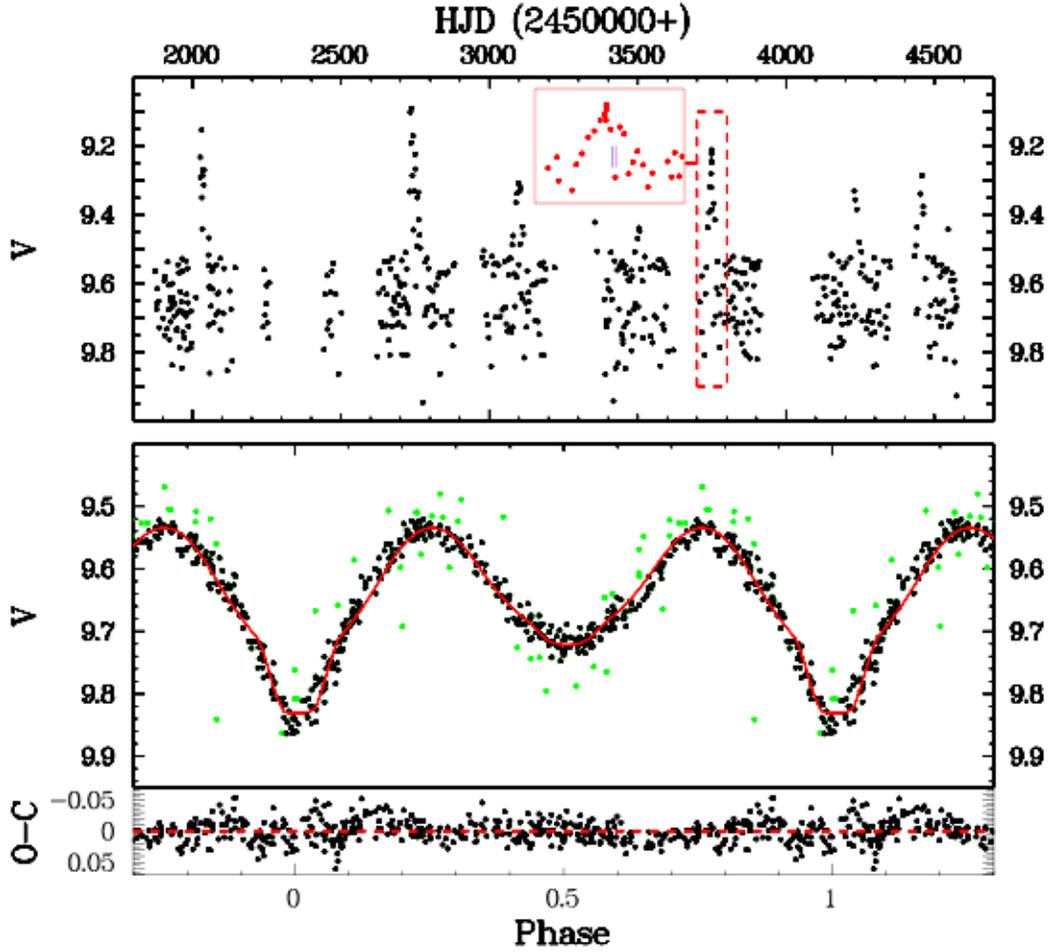}}
      \caption[]{{\em Top:} ASAS-3 long term light curve of V1129~Cen. The
                 insert contains an expanded view of the outburst selected
                 by the broken-lined box. The vertical lines mark the epochs 
                 of detection (blue) and non-detection (magenta)
                 of He~II $\lambda 4686\, {\AA}$ emission by
                 Walter et al. (2006).
                 {\em Centre:} The same data (without outbursts) folded
                 on the orbital period. The outlying green data points
                 were disregarded in the model fits discussed in
                 Sect.~\ref{V1129 Cen Model calculations}. The red graph
                 represents the best model fit.
                 {\em Bottom:} Difference between the observed light curve
                 and the best model fit. The zero level is indicated by the
                 red broken line in order to better visualize systematic
                 deviations. (For interpretation
                 of the references to colour in this figure legend, the
                 reader is referred to the web version of this article.)}
\label{v1129cen-asas}
\end{figure}

In the particular case of V1129~Cen, however, apart from variations typical for
such stars, recurring at a period of 0.893025 days, 
S. Otero\footnote{The internet links to the corresponding communications 
cited in the online version of the Ritter \& Kolb catalogue 
(http://varsao.com.ar/NSV\_19488.htm) or in Walter et al. (2006) 
(http://ar.geocities.com/varsao/NSV\_19448.htm) appear not be active any
more.} found faint outbursts with a duration of $\sim$40 days recurring 
on the time scale of one year. The ASAS (Pojmanski 2002) long term light 
curve contains several such events which reach an amplitude of up to 
$0^{\raisebox{.3ex}{\scriptsize m}}_{\raisebox{.6ex}{\hspace{.17em}.}}6$
(upper frame of Fig.~\ref{v1129cen-asas}). 
The spectral type of F2~V of V1129~Cen (Houk 1978) is
later than that of the large majority of $\beta$~Lyr stars but much earlier
than that of the donor star in any CV.
Unusual for a star of this type, Walter et al. (2006) 
observed emission of He~II $\lambda 4686\, {\AA}$ on 2006, Jan 16.2 UT
which, however, was absent on 2006, Jan 19.3 UT. Both of these observations
ocurred during an outburst as is shown by the insert in the figure, where
the corresponding epochs are marked by vertical lines. The authors leave the
question open
whether the emission was transient or if the source was eclipsed during the
second observation. It is not clear what causes this unusual (for a 
$\beta$~Lyr star) behaviour. Ritter \& Kolb (2003) have included the star as 
a possible U~Gem type dwarf nova in the on-line version of their catalogue.

\input epsf

If the system indeed contains a dwarf nova or behaves like one, persistent mass 
transfer through
an accretion disk should take place and thus flickering should be expected to
be present. Whether this would be observable or not depends on the degree of
modulation of the flickering light source and its relative contribution to
the total light of this peculiar system. In order to verify the presence
of flickering and to investigate the question whether or not the properties 
of V1129~Cen are compatible with a dwarf nova classification, I observed 
the star on several occasions in 2014, 2015 and
2016. Because of their superior quality I will concentrate here on the 2016 
light curves. These data are complemented by observations retrieved from
the ASAS data archive. Additionally, I obtained a few spectra in 2015 in order 
to verify the eventual presence of emission lines as observed by
Walter et al. (2006).

This study is organized as follows:
In Sect.~\ref{Observations and data reductions} the observations and data
reduction techniques are briefly presented. Sect.~\ref{V1129 Cen} then deals
with the results of the observations and of model calculations.
A discussion follows in Sect.~\ref{V1129 Cen Discussion}. Finally, the
conclusions are briefly summarized in Sect.~\ref{Conclusions}.

\section{Observations and data reductions}
\label{Observations and data reductions}

All photometric observations were obtained at the 0.6-m Zeiss and the 
0.6-m Boller \& 
Chivens telescopes of the Observat\'orio do Pico dos Dias (OPD), operated by 
the Laborat\'orio Nacional de Astrof\'{\i}sica, Brazil. 
Time series imaging of the field around the target star was performed
using cameras of type Andor iKon-L936-B and iKon-L936-EX2 equipped with 
back illuminated, visually optimized CCDs.
A summary of the observations is 
given in Table~\ref{Journal of observations}. Some light curves contain
gaps caused by intermittent clouds or technical reasons. In order to resolve 
any rapid flickering variations the integration
times were kept short. Together with the small readout times of the detectors
this resulted in a time resolution of the order of 
$5^{\raisebox{.3ex}{\scriptsize s}}$. In contrast to
observations of other targets within the observing project mentioned in
Sect.~\ref{Introduction}, in spite of the short integration times the high
brightness of V1129~Cen not only permitted but demanded (in order to avoid
saturation) the use of a filter. A $B$ filter was chosen. Even so, I did not
perform a rigorous photometric calibration but express the brightness as 
the magnitude difference between the target and the nearby comparison star
UCAC4 223-607051 ($B = 13.564$; Zacharias et al.\ 2013), 
the constancy of which was verified through the observations of several 
check stars\footnote{It seems that the comparison star shows small 
variations on the time scale of months and years which, however, have no
bearing on the results of this study.}. The average nightly $B$ magnitude 
of the target is included in Table~\ref{Journal of observations}. 

\begin{table}
\begin{center}

\caption{Journal of observations}
\label{Journal of observations}

\hspace{1ex}

\begin{tabular}{lccl}
\hline
Date & Start & End & \multicolumn{1}{c}{$B$} \\
     & (UT)  & (UT)&     \\
\hline
2016 Mar 09 & \phantom{2}1:19 & \phantom{2}5:10 & $^{*}$  \\
2016 Apr 05 & \phantom{2}0:20 & \phantom{2}7:11 & 10.2 \\
2016 Apr 05/06 & 23:56        & \phantom{2}5:53 & 10.1 \\
2016 Apr 06/07 & 23:56        & \phantom{2}6:30 & 10.1 \\
2016 Apr 08 & \phantom{2}1:18 & \phantom{2}6:06 & 10.2 \\
2016 Apr 14 & \phantom{2}0:31 & \phantom{2}1:51 & 10.3 \\
2016 Apr 14/15 & 23:55        & \phantom{2}5:56 & 10.1 \\ [1ex]
2015 Feb 14 & \phantom{2}5:58 & \phantom{2}7:49 & $^{**}$ \\
\hline
\multicolumn{4}{l}{$^{\phantom{*}*}$ unreliable} \\
\multicolumn{4}{l}{$^{**}$ spectroscopic observations} \\
\end{tabular}
\end{center}
\end{table}
%

In addition to the photometric observations, eight spectra of 600 sec 
exposure time were obtained on 2015, February 14, at the 1.6-m 
Perkin Elmer telescope of OPD. An Andor iKon-L936-BR-DD camera was employed. 
Exposures of a He-Ar lamp for wavelength calibration were taken after every 
second stellar exposure. From the FWHM of the lines in the comparison 
spectra a spectral resolution of $\approx$4~\AA\ is estimated.

Basic data reduction (biasing, flat-fielding) was performed using IRAF. 
For the construction of light curves aperture photometry routines 
implemented in the MIRA software system (Bruch 1993) were employed. The
same system was used for all further data reductions and calculations. 
Throughout this
paper time is expressed in UT. However, whenever observations taken in 
different nights were combined (e.g., to fold them on the orbital period)
time was transformed into barycentric Julian Date on the Barycentric 
Dynamical Time (TDB) scale using the online tool provided by 
Eastman et al. (2010) in order to take into account variations of the 
light travel time within the solar system. 

\section{Results}
\label{V1129 Cen}

\begin{figure}
   \parbox[]{0.1cm}{\epsfxsize=14cm\epsfbox{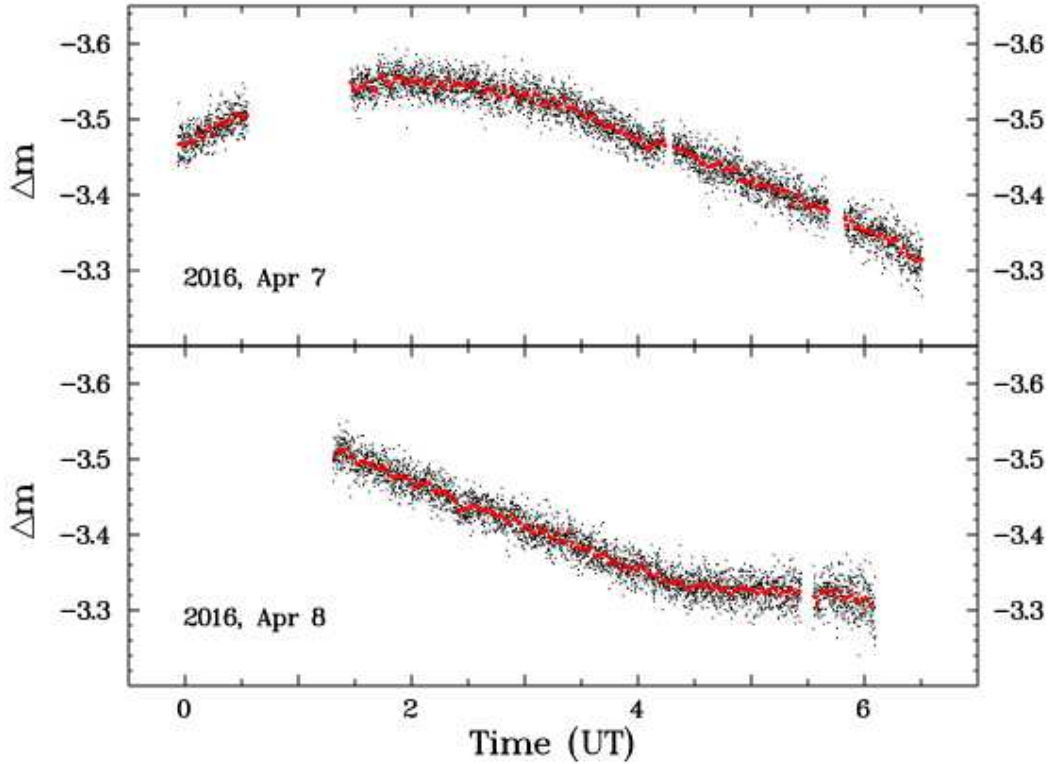}}
      \caption[]{Light curves of V1129~Cen in two nights in 2016. The
                 black dots are the original data points, while the red
                 dots represent the same data, binned in 2 minute intervals.
                 (For interpretation
                 of the references to colour in this figure legend, the
                 reader is referred to the web version of this article.)}
\label{v1129cen-lightc}
\end{figure}

The light curves are dominated by variations on hourly time scales,
reflecting the $\beta$~Lyr type variations, but contain no obvious flickering.
As examples, Fig.~\ref{v1129cen-lightc} shows two light curves of 2016,
April 7 and 8. The black dots represent the data points at the original
time resolution, while the same data, binned in intervals of 2 minutes, 
are shown in red. 

Before dealing in more detail with the issue of flickering, I first turn to 
the $\beta$~Lyr type variations of V1129~Cen. To this end, the individual
light curves were folded on the above quoted period, using as zero
point of phase the epoch of primary minimum (as cited on the AAVSO 
International Variable Star Index 
webpage\footnote{https://www.aavso.org/vsx/}).
In two nights a small magnitude adjustment was applied, calculated from the 
difference of the differential magnitudes in the respective phase intervals 
during the night in question and the other nights. This is probably due to
slight variability of the quite red 
($B-V = 1^{\raisebox{.3ex}{\scriptsize m}}_{\raisebox{.6ex}{\hspace{.17em}.}}53$; 
Zacharias et al. 2013) primary comparison star as revealed by a comparison 
with two check stars.

\begin{figure}
   \parbox[]{0.1cm}{\epsfxsize=14cm\epsfbox{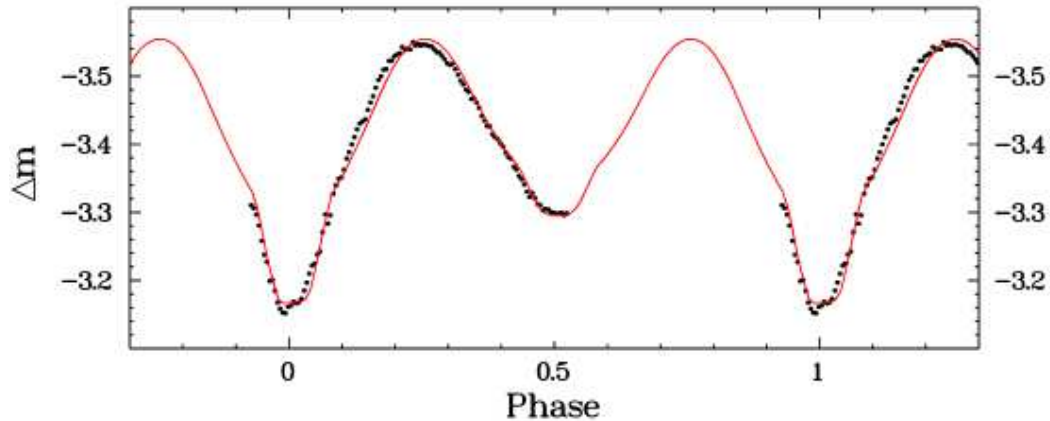}}
      \caption[]{Light curves of V1129~Cen folded on the period of the
                 $\beta$~Lyr type variations, binned in phase intervals 
                 of 0.005 (dots). The red line represents the best model 
                 fit (see Sect.~\ref{V1129 Cen Model calculations}).
                 (For interpretation
                 of the references to colour in this figure legend, the
                 reader is referred to the web version of this article.)}
\label{v1129cen-fold}
\end{figure}

The resulting $\beta$~Lyr type light curve, shown in Fig.~\ref{v1129cen-fold},
binned in phase intervals of width 0.005, does not cover all phases. Moreover,
a small shift of the primary minimum with respect to phase 0 (already 
corrected for in the figure) was observed. Its magnitude was determined to 
be $0.025 \pm 0.001$ by fitting polynomials of various degrees to the minimum.
This means that the period communicated by Otero requires a slight correction.
The observed phase shift, the minimum epoch (referring to 2002) and the 
minimum epoch observed in 2016 then permit to calculate updated ephemeries 
for V1129~Cen:
 
\begin{displaymath}
{\rm BJD_{min}} = 2457483.584 (1) + 0.8930290 (2) \times E
\end{displaymath}

{\parindent0em 
where $E$ is the cycle number. This does, or course, not take into account a
possible period variations such as that observed in the prototype star
$\beta$~Lyr (Harmanec \& Scholz 1993) at a much higher rate (19 sec/yr) that 
any possible variation implied by the difference of Otero's period and the 
present value.}

For comparison, the ASAS-3 data were also folded on the orbital period
(rejecting the observations taken during outbursts; Fig.~\ref{v1129cen-asas},
bottom). While noisy, the $\beta$~Lyr type variations are obvious. The lower
amplitude compared to Fig.~\ref{v1129cen-fold} may be due to the different
passband of the ASAS-3 data ($V$ vs. $B$).

\subsection{Spectrum}
\label{V1129 Cen Spectrum}

\begin{figure}
   \parbox[]{0.1cm}{\epsfxsize=14cm\epsfbox{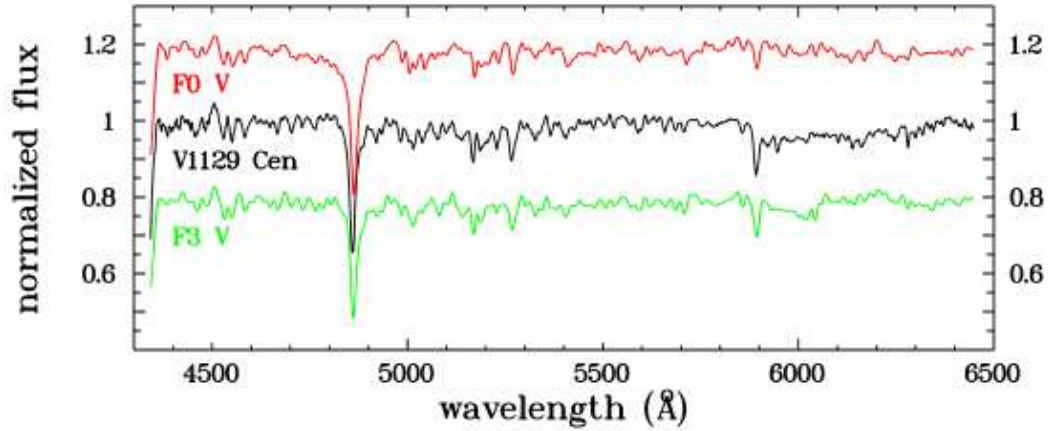}}
      \caption[]{Average continuum normalized spectrum of V1129~Cen on 
                 2015, Feb. 14 (black), together with the spectra of two
                 standard stars of similar spectral type F0~V (red) and
                 F3~V (green). (For interpretation
                 of the references to colour in this figure legend, the
                 reader is referred to the web version of this article.)}
\label{v1129cen-spec}
\end{figure}

Having in mind the report of Walter et al.\ (2006) of transient He~II 
$\lambda 4686\, {\AA}$ emission in the spectrum of V1129~Cen, I obtained the
spectroscopic observations mentioned in 
Sect.~\ref{Observations and data reductions}.
The mean of eight individual exposures is shown in Fig.~\ref{v1129cen-spec}
(black curve). Since no flux calibration of the spectra was performed it
is shown here normalized to the continuum. For comparison, standard star 
spectra of spectral type F0~V and F3~V (i.e., close to the spectral type of 
V1129~Cen), taken from the compilation of Jacoby et al.\ (1984) and normalized
in the same way are also shown in the figure (shifted upward and downward
for clarity). Their resolution was degraded to  match that of V1129~Cen.
No trace of $\lambda 4686\, {\AA}$ emission is seen. 

\subsection{Flickering}
\label{V1129 Cen Flickering}

I turn my attention now to the implications of the absence of detectable
flickering in the light curves of V1129~Cen. 

I first determine the scatter of the data points of the binned versions of
the light curves shown in Fig.~\ref{v1129cen-lightc} (adding also the night 
of 2016, April 6) after subtraction of the orbital variations. To this end a 
Gaussian was fit to the distribution of the difference between data points of
the binned light curves and a Fourier filtered version of the same data which 
removes variations on time scales 
$>$$30^{\raisebox{.3ex}{\scriptsize m}}$. In all nights it has a FWHM of about 
$\delta m_0 = 0^{\raisebox{.3ex}{\scriptsize m}}_{\raisebox{.6ex}{\hspace{.17em}.}}01$.
 
\begin{figure}
   \parbox[]{0.1cm}{\epsfxsize=14cm\epsfbox{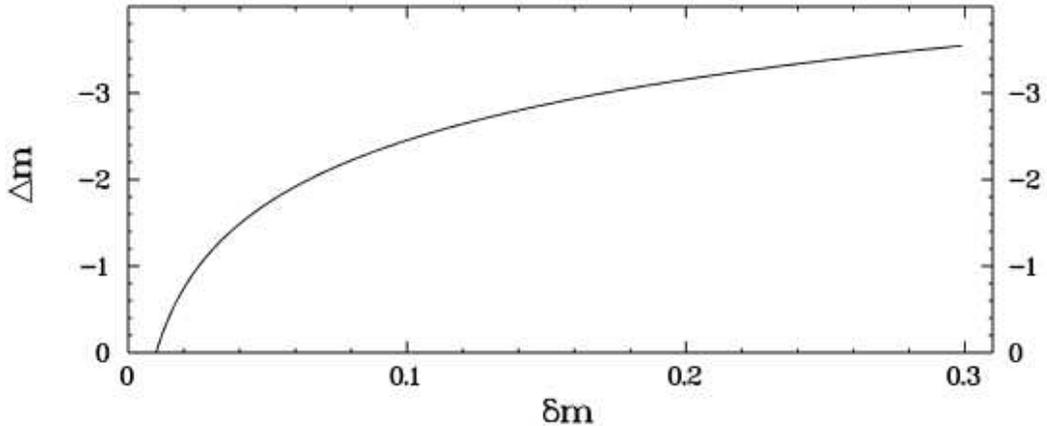}}
      \caption[]{Magnitude difference $\Delta m$ between a light source with
                 a flickering amplitude $\delta m$ and the same plus an
                 additional (constant) light source such that the observed
                 flickering amplitude drops to $\delta m_0 = 0.01$.}
\label{v1129cen-flick}
\end{figure}

What must be the magnitude difference between a flickering light source
and a brighter constant star in order to render the flickering unobservable?
Assuming the presence of a light source in the system which flickers 
such that a light curve treated in the same way as above leads to a distribution
of data points with a FWHM of $\delta m$ it is possible to calculate as a
function of $\delta m$ the magnitude difference $\Delta m$ of that light 
source and of the entire system necessary for the observed FWHM not 
to exceed $\delta m_0$. This leads to the relationship shown in 
Fig.~\ref{v1129cen-flick}. Considering that the total amplitude of the
flickering variations is significantly larger than the FWHM of the 
distribution of data points [e.g., in V504~Cen Bruch (2017b)
observed a total amplitude of 
$0^{\raisebox{.3ex}{\scriptsize m}}_{\raisebox{.6ex}{\hspace{.17em}.}}62$, 
while the FWHM does not exceed 
$0^{\raisebox{.3ex}{\scriptsize m}}_{\raisebox{.6ex}{\hspace{.17em}.}}16$] the range of 
$\delta m$ in the figure extends to extremely strong flickering.


According to Gaia DR1 (Brown et al. 2016), 
V1129~Cen has a parallax of $3.21 \pm 0.40$~mas.
This translates into a distance of $312 \pm 39$~pc. The ASAS-3 long term 
light curve shows that the $V$ magnitude varies between 
$9^{\raisebox{.3ex}{\scriptsize m}}_{\raisebox{.6ex}{\hspace{.17em}.}}47$ and
$9^{\raisebox{.3ex}{\scriptsize m}}_{\raisebox{.6ex}{\hspace{.17em}.}}96$
(disregarding outburst), with an average of 
$9^{\raisebox{.3ex}{\scriptsize m}}_{\raisebox{.6ex}{\hspace{.17em}.}}65$.
While the transformation of ASAS magnitudes to a standard photometric system 
may not be particularly accurate, errors are expected not to exceed a typical
value of 
$0^{\raisebox{.3ex}{\scriptsize m}}_{\raisebox{.6ex}{\hspace{.17em}.}}05$\footnote{http://www.astrouw.edu.pl/~gp/asas/explanations.html}. 
Moreover, the average magnitude of
$9^{\raisebox{.3ex}{\scriptsize m}}_{\raisebox{.6ex}{\hspace{.17em}.}}65$ 
is identical to the $V$ magnitude cited in the Tycho-2 catalogue 
(H{\o}g et al. 2000).

The interstellar absorption towards V1129 Cen appears to be small. The
observed colours 
{$B-V=0.38$; $U-B=0$; Kilkenny \& Laing 1990)\footnote{However,
the authors marked these values as uncertain.} match quite well
with those of an unreddened star of the same spectral type 
{F2~V; $B-V = 0.35$, $U-B = 0.00$; FitzGerald 1970}.
Neglecting thus absorption, the distance
and the apparent average magnitude translate into an absolute magnitude
of $M_V = 2.18 \pm 0.29$, slightly 
($\sim$$0^{\raisebox{.3ex}{\scriptsize m}}_{\raisebox{.6ex}{\hspace{.17em}.}}75$)
brighter than an 
F2 type main sequence star (from interpolation in the 
tables of Allen 1973). This difference cannot be explained by the
contribution of the binary companion to the total system light since model 
calculations (Sect.~\ref{V1129 Cen Model calculations}) show this contribution
to be much less than that of the F2 star. The difference rather indicates
that the latter has slightly evolved off the main sequence.

The quiescent magnitudes of dwarf novae encompass a wide range. Fig.~3.5 of
Warner (1995) suggests that ordinary U~Gem stars have a brightness
fainter than 
$M_V = 7^{\raisebox{.3ex}{\scriptsize m}}_{\raisebox{.6ex}{\hspace{.17em}.}}2.$
While in long period
systems it includes a non-negligible contribution of the mass donor, for
the sake of a conservative upper limit I consider this value to be the 
magnitude of a possible accretion disk in V1129~Cen. The magnitude difference
between the entire system and the accretion disk is thus at least $\Delta m = 
-5^{\raisebox{.3ex}{\scriptsize m}}_{\raisebox{.6ex}{\hspace{.17em}.}}0$.
Comparing this value with the graph in 
Fig.~\ref{v1129cen-flick} it is obvious that the not flickering light
sources in V1129~Cen can easily hide any flickering even if the disk light
would be 100\% modulated. 

\subsection{Model calculations}
\label{V1129 Cen Model calculations}

The complete phase coverage of the ASAS data warrents an attempt to
model the light curve of V1129~Cen in the expectation that some system
parameters can be delimited. To this end, I employ the Wilson-Devinney
code (Wilson \& Devinney 1971, Wilson 1979) as implemented in MIRA.
Before proceeding, a word on nomenclature is in order to avoid
confusion with nomenclature usually used in CV research: I will refer to
the optically dominating F2 star as the primary component, independent whether
it is the mass gainer or mass loser (if there is mass transfer in the system)
or whether it is the more massive or the less massive star. It will be 
designated by the index 1 subsequently. Consequently its companion is the 
secondary star (index 2). The mass ratio is defined as $q = M_2/M_1$.

Considering the large number of model parameters required by the 
Wilson-Devinney code to calculate
a light curve it is appropriate to fix as many of
them as possible before trying to adjust the model light curve  
to the observed data. 

It turns out that the atmospheric parameters
albedo $\cal{A}$, limb darkening coefficient $u$ and gravity darkening
coefficient $y$ have only a minor influence on the results. Therefore, 
$u$\footnote{A simple linear law of the kind $
I(\mu)/I(1) = 1 - u (1-\mu)$ is used. Here, $I(1)$ is the
specific intensity at the centre of the stellar disk, and $\mu = \cos \gamma$,
where $\gamma$ is the angle between the line of sight and the emergent
radiation.} and $y$ are interpolated in the tables of Claret \& Bloemen (2011)
[using their results based on ATLAS model, least squares calculations,
adopting solar chemical composition, the surface gravity of a normal F2 
main sequence star and
no microturbulence; for details, see Claret \& Bloemen (2011)]
at the temperature of 7\,040~K as determined by Kordopatis et al.\ (2013)
for the primary star. Similarly, $u$ and $y$  for the secondary component
refer to a temperature of 4\,500~K (determined from preliminary model fits) 
and a surface gravity calculated from the mass and radius of a main sequence 
star of that temperature. According
to Rafert \& Twigg (1980) hotter stars with radiative envelopes should have
an albedo of $\cal{A}$$ = 1.0$ while for cooler stars with convective envelopes
$\cal{A}$$ = 0.5$. I adopt the latter value for the secondary star of V1129~Cen.
The temperature of the primary falls in the transition region between the
two regimes. For simplicity, I adopt $\cal{A}$$_1 = 0.75$. 
Furthermore, the primary temperature is fixed to the above mentioned value
of $T_1 = 7\,040$~K. A phase shift to make up for a possible
slight error of the epoch of primary eclipse was fixed to the value determined
in preliminary calculations. Any contribution of an accretion disk and/or
hot spots possibly present in the system was ignored.

The model parameters left free to be adjusted to the data where then 
the mass ratio $q$ of the components, the orbital inclination $i$, 
the temperature $T_2$  and the dimensionless surface potential $\Omega_2$ 
of the secondary star. $\beta$~Lyr stars are binaries in a tight orbit, but
it is not always evident if they are detached or semi-detached. Therefore,
calculations for both cases were performed, choosing the corresponding mode
of the Wilson-Devinney code. In the latter case the surface potential of 
the primary star is determined by the mass ratio which defines the potential 
at the Roche surface. In the alternative case no limitations 
on the size of the components relative to their Roche lobes is assumed and 
the surface potential of the primary was also left free to be adjusted.
Finally, the normalization constant was also considered a free parameter. 
It turned out that the best fit parameters of the detached model were not
significantly different from those of the semi-detached model. Therefore,
to be definite, I will subsequently only regard the results derived from
the latter.

The SIMPLEX algorithm (Caceci \& Cacheris 1994) was adopted to find the optimal
model parameters which lead to the minimal $\chi^2$ between observations
and calculations. Some outlying data points in the observed light curve
(green dots in Fig.~\ref{v1129cen-asas}) were disregarded. 
The best fit model is shown in the central frame of Fig.~\ref{v1129cen-asas} 
as a red curve. The lower frame contains the differences between the observed 
and calculated data. The broken red line indicates the zero level in
order to better visualize systematic deviations of the $O-C$ curve from zero.
The fit parameters are summarized in Table~\ref{V1129 Cen fit parameters}. 
Here, the Roche lobe filling factor of the secondary is calculated from its
surface potential and is thus not an independent quantity.

\begin{table}

\caption{Parameters of the model fit to the V1129~Cen light curve}
\label{V1129 Cen fit parameters}

\hspace{1ex}

\begin{tabular}{lllrrr}
\hline
Parameter &  &  & \multicolumn{1}{c}{Best fit}  &
\multicolumn{1}{c}{Best fit}   & \multicolumn{1}{c}{acceptable range} \\
          &        &  & \multicolumn{1}{c}{($V$ band)} & 
\multicolumn{1}{c}{($B$ band)} & \multicolumn{1}{c}{(from $V$ band data)} \\
\hline
Albedo (prim.)                        & $ \cal{A}$$_1$ & 
fixed    & 0.75     & 0.75            &   \\
Albedo (sec.)                         & $ \cal{A}$$_2$ &
fixed    & 0.50     & 0.50            &   \\
limb dark. coef. (pr.)                & $u_1$          & 
fixed    & 0.55     & 0.65            &   \\
limb dark. coef. (sec.)               & $u_2$          & 
fixed    & 0.79     & 0.89            &   \\
grav. dark. coef. (pr.)               & $y_1$          & 
fixed    & 0.24     & 0.29            &   \\
grav. dark. coef. (sec.)              & $y_2$          & 
fixed    & 0.64     & 0.83            &   \\
Temperature (prim.) (K)               & $T_1$          & 
fixed    & 7040\phantom{.00} &7040\phantom{.00}          &   \\ 
Temperature (sec.)\hspace{1ex} (K)    & $T_2$          & 
adjusted & 4490\phantom{.00} & 5210\phantom{.00}         & 
($<$4050\phantom{.00} \ldots \phantom{$>$}6550\phantom{.00}) \\ 
Orbital inclination\hspace{1ex} ($^{\raisebox{.3ex}{\scriptsize o}}$) & $i$      & 
adjusted & 74.6\phantom{0} & 74.3\phantom{0} & 
(\phantom{$<$45}67\phantom{.00} \ldots \phantom{$>$65}90\phantom{.00}) \\  
Mass ratio                            & $q$            
& adjusted & 0.61    & 0.61           & 
(\phantom{$<$650}0.45 \ldots \phantom{$>$655}0.77) \\
Surface potential (sec.)              & $\Omega_2$     & 
adjusted & 7.55   & 7.32              & 
(\phantom{$<$655}5.8\phantom{0} \dots \phantom{65}$>$11\phantom{.00}) \\
\multicolumn{2}{l}{Roche lobe filling factor (sec.)}                     & 
adjusted & 0.27   & 0.28              & 
(\phantom{$<$655}0.17 \ldots \phantom{$>$655}0.37) \\ 
\hline
\end{tabular}
\end{table}
%

The model fit is not completely satisfactory. There are systematic residuals
between data and fit. In particular, the fit appears to slightly underestimate
the brightness after the primary minimum (phase range $0.1 < \phi < 0.2$ and
overestimates it after the secondary minimum ($0.6 < \phi < 0.7)$. The formal
deficiency of the fit is also evident from the elevated value of the
reduced $\chi^2_{\rm r,min} = 2.9$. Some ingredients are therefore probably 
missing in the model. If the outbursts of V1129~Cen are indeed related to
dwarf nova outbursts this is not surprising because an
accretion disk (and possibly associated bright spots) are then expected
to be present in the system. These cannot be modeled by the 
Wilson-Devinney code.

\begin{figure}
   \parbox[]{0.1cm}{\epsfxsize=14cm\epsfbox{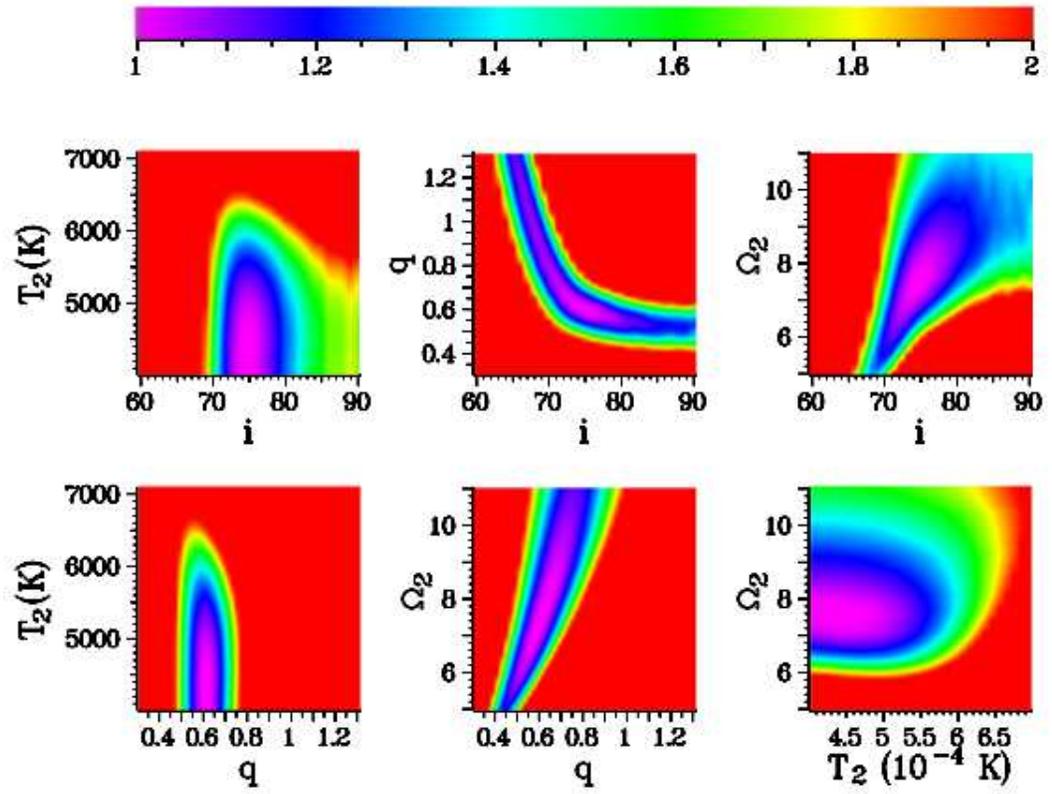}}
      \caption[]{Two-dimensional cuts through the $\chi^2_{\rm r}$
                 hyperspace, defined by the residuals between the observed
                 light curve of V1129~Cen and model calculations, at the 
                 location of the best fit parameters.
                 The colour coding (see colour bar at the top of the figure)
                 is such that purple corresponds to 
                 $\chi^2_{\rm r} = \chi^2_{\rm r,min}$ and dark red
                 represents $\chi^2_{\rm r} \ge 2 \chi^2_{\rm r,min}$.
                 (For interpretation
                 of the references to colour in this figure legend, the
                 reader is referred to the web version of this article.)}    
\label{v1129cen-parerr}
\end{figure}

Even so, considering that the orbital modulation is evidently dominated
the the ellipsoidal variations of the primary component together with a 
substantial primary and a smaller secondary eclipse (all determined by
the component temperatures, their relative sizes and the mass ratio), the 
missing model ingredients may demand small corrections, but the best 
fit parameters should at least approximately reflect reality.

Parameter correlations make it difficult to assign meaningful statistical
errors to the parameter values. In order to investigate this issue
Fig.~\ref{v1129cen-parerr} shows two-dimensional cuts through the 
$\chi^2_{\rm r}$ hyperspace at the location of the best fit parameters. 
In order to facilitate comparison, the colours coding of all frames is 
such that purple corresponds the $\chi^2_{\rm r,min}$ and
dark red to twice that value or higher (see colour bar at the top of the
figure). It is then seen that, for instance, the orbital inclination and 
the mass ratio are strongly correlated (central frame in the upper row
of Fig.~\ref{v1129cen-parerr}) which makes it impossible to determine either
of them with any degree of precision. Assuming as criterion that solutions
leading to $\chi^2_{\rm r} > 2 \chi^2_{\rm r,min}$ are unacceptable the diagram
shows that the mass ratio can be anything from 0.39 up to a value beyond the 
limits of 
the explored parameter range. However, other cuts through the $\chi^2_{\rm r}$
hyperspace permit to better restrict $q$ (i.e., the $q - T_2$ plane; lower
left frame of Fig.~\ref{v1129cen-parerr}). Exploring the individual cuts in
this way leads to permitted parameter ranges as quoted in the last columns of  
Table~\ref{V1129 Cen fit parameters}\footnote{The lower limit for the range
of $T_2$ is ill defined because the Wilson-Devinney code issued warnings when
$T_2 < 4050$~K. The corresponding model calculations were then ignored.}

As a check, the Wilson Devinney model was also fit to the $B$ light curve
(red line in Fig.~\ref{v1129cen-fold}). Again, the fit is not perfect,
exhibiting the same excess of observed light in the phase range after
primary minium already seen in the V-band data. With the exception of
the secondary star temperature which is higher by $\approx$700~K when
the $B$ band data are used, the best fit parameters listed in 
Table~\ref{V1129 Cen fit parameters} are practically identical to those
derived from the $V$ band. Even so, $T_2$ remains comfortably within the
acceptable range. The agreement of the results obtained from data in 
different bands and using radically different observing procedures gives
confidence that they are not corrupted by errors or systematics
of the observations.

The results of the model calculations nicely fit in with independent 
knowledge about V1129~Cen. Assuming the mass $\cal{M}$$_1$ of the primary not 
to be significantly different from that of a 
normal F2~V star ($\sim$1.55~$\cal{M}_\odot$; Allen 1973) the 
mass ratio and the orbital period together with Kepler's third
law yield the component separation $A$. Using the approximation for the volume
radius of the Roche lobe provided by Eggleton (1983), the assumption
of a semi-detached configuration determines the radius $R_1$ of the primary 
in units of $A$ as a function of $q$. For the best fit value of $q$ this
results in a primary star radius of $R_1 = 2.2\ R_\odot$. This is 1.7 times the
radius of a F0~V star according to Allen 1973),
confirming the conclusion drawn in Sect.~\ref{V1129 Cen Flickering} that
the star has evolved off the main sequence. Since the brightness scales with
the square of the radius, the V1129~Cen primary should be 
$1^{\raisebox{.3ex}{\scriptsize m}}_{\raisebox{.6ex}{\hspace{.17em}.}}2$
brighter that its main sequence equivalent. This is somewhat more than the
values found in Sect.~\ref{V1129 Cen Flickering} 
($0^{\raisebox{.3ex}{\scriptsize m}}_{\raisebox{.6ex}{\hspace{.17em}.}}75$) but not
significantly so considering the uncertainty of $M_V$ and $q$.

\section{Discussion}
\label{V1129 Cen Discussion}

The main issue concerning V1129~Cen is the question about the nature of the
semi-periodic outbursts. Are these genuine dwarf nova type eruptions? If
so, how can they come about in a system that otherwise appears to have a
configuration different from normal cataclysmic variables?

Outbursts of dwarf nova are caused by an increase of matter transferred
through an accretion disk and the corresponding release of energy. This may 
be due to a limit cycle in the disk which during quiescence is in a cool and
low viscosity state. Matter transferred from a 
donor star increases the disk mass and its temperature until the temperature
for hydrogen ionization is reached, resulting in an increased viscosity
which causes the disk matter to be dumped on the central star 
(the thermal-viscous instability model; Lasota 2001).
While this is the commonly adopted mechanism for dwarf nova outbursts, 
at least in some systems it seems not to work 
and there is evidence that instead an increased mass transfer from the
companion star is responsible for the brightening of the accretions disk and
thus the dwarf nova outburst [see the discussion in Baptista (2012)]. 
 
\subsection{Outburst characteristics}
\label{Outburst characteristics}

What are the characteristics of the brightenings of V1129~Cen and how do they
relate to those of normal dwarf nova outbursts?
They last for about 40 days which is significantly longer than 
observed in most dwarf novae. However, as Szkody \& Mattei (1984) and
Gicger (1987) showed, there is a clear correlation between orbital
period and outburst duration. Extrapolating the relation of 
Gicger (1987) to the period of V1129~Cen leads to 40.7 days in
remarkable agreement with the average outburst duration measured in the
ASAS-3 light curve. 

While most dwarf nova outbursts rise rapidly and 
decline more slowly, the V1129~Cen outbursts are more symmetrical. This
may also be a consequence of the long orbital period since outburst of
other long period CVs show similar shapes [BV~Cen, $P_{\rm orb}=0.611$ days, 
Bateson (1974); GK~Per\footnote{GK~Per is well known to be a 
classical nova, but it also exhibits dwarf nova outbursts.}, 
$P_{\rm orb}=1.997$ days, Pezzuto et al.\ (1996), Evans et al.\ (2009); 
V630~Cas, $P_{\rm orb}=2.564$ days, Shears \& Poyner (2009)].

Similarly, using a mean outburst interval of 347 days (from
Fig.~\ref{v1129cen-asas}, assuming one unobserved outburst close to
JD~2452400) V1129~Cen fits in very nicely in a linear relationship
between the orbital period and the logarithm of the outburst intervals of 
BV~Cen {interval: 150 days; Menzies et al. 1986}, GK~Per ($\sim$ 1060 days; 
deduced from the AAVSO long term light curve, assuming some outbursts to 
have been missed) and V630~Cas (17 years; Shears \& Poyner 2019). 

While not
irrefutable proof, the similarity of the outburst properties of V1129~Cen
with those of long period dwarf novae suggest that the nature of the eruptions
is similar.

\subsection{Scenarios}
\label{Scenarios}

If the bright states in V1129~Cen are in fact dwarf nova type 
outbursts there must be an accretion disk somewhere in the system. In 
principle, at least three possible
scenarios can be envisaged: (1) The source of the outbursts is accidentally
in the line of sight to V1129~Cen but not physically related to it; (2) 
V1129~Cen is not a simple binary star but a quadruple formed by two pairs,
i.e., the dominating $\beta$~Lyr type component which has a normal dwarf nova 
as a companion at a distance where the evolution of either system does not
interfere with the other one;
(3) the mass gainer in V1129~Cen is surrounded by an accretion disk as
modeled, e.g., in the case of the prototype $\beta$~Lyr by 
Mennickent \& Djura\v{s}evi\'c (2013).
The first possibility can, or course, not be excluded, but it is quite
unlikely considering the low space density of dwarf novae. Therefore, I
will not consider it further. 

\subsubsection{A hierarchical quadruple system}
\label{A hierarchical quadruple system}

Exploring the second scenario, I first remark that the considerations of
Sect.~\ref{Outburst characteristics} about correlations between the 
outburst duration, shape and intervals, and the orbital period are 
irrelevant in this case because the observed period of V1129~Cen is
then not that of the dwarf nova. The question may be asked whether
this scenario is compatible with basic evolutionary
considerations. Are the evolutionary time scales of the $\beta$~Lyr star and
the dwarf nova compatible with their co-existence in a single multiple star 
system? This comes down to the question if the progenitor of the white dwarf
in the dwarf nova system was more massive than the F2 star because in
that case the former had enough time to go through a common envelope phase
and become a cataclysmic variable while the F star still remains on or close
to the main sequence. 

Several initial -- final mass relations for white dwarfs have been published
in the literature. Let $\cal{M}$$_{\rm prog}$ be the mass of the progenitor
of a white dwarf of mass $\cal{M}$$_{\rm WD}$. Using any of the relations
given by Zhao et al.\ (2012), Salaris et al.\ (2009) or Catalan et al.\ (2008),
the requirement that $\cal{M}$$_{\rm prog} > 1.55$$\cal{M}$$_\odot$ 
[i.e., the mass of a F2~V star according to Allen (1973)] leads
to $\cal{M}$$_{\rm WD} > 0.60$$\cal{M}$$_\odot$. This holds for single stars.
As Ritter (2010) points out, in binaries the mass transfer sets a
premature end to the nuclear evolution of the donor star. Therefore, the
resulting white dwarf mass is smaller than in the case of single star 
evolution. The mean mass of white dwarfs in CVs is 0.83~$\cal{M}$$_\odot$ 
(Zorotovic et al. 2011). Thus, there is ample
space for the progenitor to have a mass high enough to 
evolve into a red giant and to initiate the common envelope phase which
results in the formation of a CV before the main component, i.e., the F star,
leaves the main sequence.

On the other hand, the observed outburst amplitude, while not rendering
this scenario impossible, casts some doubt upon it. Fig.~\ref{v1129cen-asas}
shows that the amplitude of different outbursts range between 
$0^{\raisebox{.3ex}{\scriptsize m}}_{\raisebox{.6ex}{\hspace{.17em}.}}6$ and
$0^{\raisebox{.3ex}{\scriptsize m}}_{\raisebox{.6ex}{\hspace{.17em}.}}4$.
To be definite, the mean of the extremes, 
$0^{\raisebox{.3ex}{\scriptsize m}}_{\raisebox{.6ex}{\hspace{.17em}.}}5$,
will be adopted here. 
Remembering that the absolute magnitude of the dominating F2 star in 
V1129~Cen is 
$M_V = 2^{\raisebox{.3ex}{\scriptsize m}}_{\raisebox{.6ex}{\hspace{.17em}.}}18$
(see Sect.~\ref{V1129 Cen Flickering})
and assuming that the secondary of the $\beta$~Lyr type system and the
quiescent dwarf nova contribute negligibly to the total light, the
magnitude of the outbursting light source should then be 
$\sim$$0^{\raisebox{.3ex}{\scriptsize m}}_{\raisebox{.6ex}{\hspace{.17em}.}}6$
fainter than the F star. Thus, its absolute magnitude is 
$\sim$$2^{\raisebox{.3ex}{\scriptsize m}}_{\raisebox{.6ex}{\hspace{.17em}.}}8$.
Together with the conservative
limit for the magnitude difference between the entire system and the
accretion disk derived in Sec.~\ref{V1129 Cen Flickering} this means that
the dwarf nova should have an outburst amplitude of at least 
$-4^{\raisebox{.3ex}{\scriptsize m}}_{\raisebox{.6ex}{\hspace{.17em}.}}4$.
This is an uncomfortably high value. Amplitudes as large as this are more 
typical for superoutbursts of SU~UMa stars than for normal dwarf nova outbursts.
But the ASAS long term light curves does not show evidence of the dichotomy
between normal and superoutbursts characterizing those systems. 
Moreover, the absolute magnitude of the outbursting light source must also
be quite high in this scenario. In normal dwarf novae the absolute $V$ band 
outburst magnitude $M_{\rm max}$ increases with the orbital period $P_{\rm orb}$,
reflecting the larger size of the accretion disk in systems with longer periods.
The relationship given in Eq.~13 of Warner(1987) results in a range
of $4.96 \ge M_{\rm max} \ge 3.15$ for $P_{\rm orb}$ between 1.5~h and 10~h. 
This holds for an average inclination of
$57^{\raisebox{.3ex}{\scriptsize o}}_{\raisebox{.6ex}{\hspace{.17em}.}}7$ of the accretion
disk. Assuming $i=74.7$ (Sect.~\ref{V1129 Cen Model calculations}) and the
inclination correction of Paczy\'nski \& Schwarzenberg-Czerny (1980) 
the disk is expected to be even
$1^{\raisebox{.3ex}{\scriptsize m}}_{\raisebox{.6ex}{\hspace{.17em}.}}08$ less luminous and
thus much fainter than the lower brightness limit estimated above. Therefore,
the dwarf nova
companion to the $\beta$~Lyr type binary in the V1129~Cen must have rather
extreme properties compared to an average dwarf nova for this scenario to be
viable.

\subsubsection{Outbursts within the $\beta$~Lyr type binary}
\label{Outbursts within the beta Lyr type binary}

Turning to the third scenario, I assume that V1129~Cen consists of only two
stellar components, one of which is surrounded by an accretion disk. In fact,
similar models have successfully been adjusted to the light curves of
several $\beta$~Lyr type systems: AU~Mon (Djura\v{s}evi\'c et al. 2010), 
V393~Cen (Mennickent et al. 2012), V455~Cyg (Djura\v{s}evi\'c et al. 2012), 
OGLE 05155332-6925581 (Garrido et al. 2013) and the prototype $\beta$~Lyr 
itself (Mennickent \& Djura\v{s}evi\'c 2013).
In all of these the disk revolves around the optically dominating primary 
star which is thus the mass gainer, receiving matter from a Roche lobe filling 
secondary star of lower mass. However, in the quoted examples
the binary is always much hotter and more massive than in the 
case of V1129~Cen, harbouring primary stars of spectral type O and B and
masses ranging from $7M_\odot$ to $13M_\odot$. Moreover the accretion disks
are all extremely massive, geometrically and optically thick, and hot. The
usual disk instability type mechanism for dwarf nova outbursts cannot work in
such disks.

But what about the alternative case where the optically dominating star is
the mass donor and the accretion disk revolves around the companion? This
configuration would be similar to that of a normal cataclysmic variable
with the difference that the donor would have a much earlier spectral type
than any other CV. 

In order to evaluate the consequences of this picture
I regard the results of the model calculations of 
Sect.~\ref{V1129 Cen Model calculations}. Although they provide 
formal values for $\Omega_2$ and $T_2$, these cannot 
be used to draw conclusions on the nature of the secondary star (here:
the mass gainer).
The Wilson-Devinney code assumes the secondary to be a spherical object 
(distorted by the Roche potential). If the companion to the F2 star is in 
reality a star surrounded by an accretion disk, the fit parameters referring to
the secondary will therefore represent an ill defined mixture of stellar and
accretion disk parameters. 

However, this does not affect the brightness of the components. Adopting the 
best fit parameters the model calculations show that the primary component 
(here: the mass looser) is $\sim$160 times brighter than the secondary 
at phase 0.25. This corresponds to a magnitude difference of 
$-5^{\raisebox{.3ex}{\scriptsize m}}_{\raisebox{.6ex}{\hspace{.17em}.}}5$, compatible with 
the minimum magnitude difference derived from the absence of obserable
flickering (see Sect.~\ref{V1129 Cen Flickering}). 
The stark brightness contrast between the components also demands a high
S/N ratio in order to detect the contribution of the expected emission lines 
from the accretion disk in the spectrum (provided that the disk is in a low
vicosity state; emission lines in the bright, high viscosity state tend to be 
weak or even replaced by absorptions). Measuring the ratio of the flux at
the top of the H$\beta$ emission line to the flux of the surrounding
continuum in the spectra of CVs reproduced by Zwitter \& Munari (1995, 1996) 
yields a maximum of $\sim$5. Taking this as an
upper limit for the corresponding ratio in the supposed accretion disk in
V1129~Cen, a S/N ratio of at least 32 is then required for a spectrum to
exhibit a trace of a H$\beta$ emission. Thus, the absence of an emission core
in the H$\beta$ absorption line in Fig.~\ref{v1129cen-spec} is not incompatible
with the idea of a dwarf nova-like accretion disk around the secondary
component.

While the large outburst amplitude derived in 
Sect.~\ref{A hierarchical quadruple system} may still represent a certain
problem, this is different for the absolute magnitude of
$\sim$$2^{\raisebox{.3ex}{\scriptsize m}}_{\raisebox{.6ex}{\hspace{.17em}.}}8$
of the outbursting light source. Provided that an extrapolation of Eq.~13 of 
Warner (1987) [Warner (1995) restricts its validity to 
orbital periods $\le$$15^{\raisebox{.3ex}{\scriptsize h}}$] to the period
of V1129~Cen ($21^{\raisebox{.3ex}{\scriptsize h}}_{\raisebox{.6ex}{\hspace{.17em}.}}4$)
does not lead to an excessive error, the accretion disk in outburst may
be even as bright as 
$1^{\raisebox{.3ex}{\scriptsize m}}_{\raisebox{.6ex}{\hspace{.17em}.}}3$.

The fact that in this scenario the mass donor is of significantly earlier
spectral type than in any other known CV does not invalidate it. In recent 
population synthesis calculations for CVs, Goliasch \& Nelson (2015) explicitly
took into account the nuclear evolution of high mass donor stars. They show
that CVs with donor star masses corresponding to early F stars can form,
in particular if the donor has already evolved off the main sequence.
At first glance, the fact that all values of $q$ within the acceptable 
range quoted in Tab.~\ref{V1129 Cen fit parameters} lead to a donor star 
mass significantly in excess of that of the mass gainer appears problematic
since this is in contrast to normal CVs where the mass donor is always 
less massive than the mass gainer. But also in this case the calculations of 
Goliasch \& Nelson (2015) indicate that the donor can be much more massive than
the gainer, in particular if it is evolved. However, the stability and the
rate of mass transfer may then become an issue. Can it be kept low enough
for the disk to remain in a quiescent low state prone to dwarf nova outbursts?
Using the same stellar evolution code as Goliasch \& Nelson (2015), 
Kalomeni et al.\ (2016) have calculated a dense grid of evolutionary tracks
for binaries with white dwarf primaries. In their Fig.~16 they plot as a
function of the donor star mass and the orbital period the ratio
$\dot{\cal M}/\dot{\cal M}_{\rm crit}$ of the mass transfer rate $\dot{\cal M}$
and the critical transfer rate $\dot{\cal M}_{\rm crit}$ above which
the accretion disk is stable against the thermal-viscous instability. The
latter is based on the stability criterion of Lasota (2001). The
figure shows that at the orbital period of V1129~Cen and a donor star mass
close the that of an unevolved (or only slightly evolved) early F type star
configurations with $\dot{\cal M}/\dot{\cal M}_{\rm crit} < 1$ occur. Thus,
dwarf nova type outbursts are possible. 

%

However, a problem arises from the large dimensions of the system which
implies a high critical disk mass transfer rate $\dot{\cal M}_{\rm crit}$ 
for outbursts to occur due a thermal-viscous disk instability and, in 
consequence, a high outburst luminosity. Smak (1983) provides an 
expression for $\dot{\cal M}_{\rm crit}$. I neglect the small correction
factor involving the the ratio between the white dwarf radius and the disk
radius $R_{\rm d}$ and follow Osaki (1996) adopting the expression
$\log{T_{\rm eff,crit}} = 3.9 - 0.1 \log{(R_{\rm d}/10^{10} {\rm cm})}$ for the 
critical disk temperature and $R_{\rm d} = 0.35 A$, where $A$ is the component
separation. $A$ is calculated from Kepler's third law, using the typical
mass of an F0~V star and the mass ratio as quoted in 
Table~\ref{V1129 Cen fit parameters}. The critical mass transfer rate for
a disk instability to occur is then $\sim$$1.8 \times 10^{-7} \cal{M}_\odot/$y.
This leads to an approximate lower limit of the outbursting disk luminosity
of $L_{\rm d,o} = G \dot{\cal{M}} \cal{M}_{\rm WD}/$$R_{\rm WD} = 637\, L_\odot$
where $G$ is the gravitational constant and the white dwarf radius $R_{\rm WD}$
has been calculated from its mass and the mass-radius relation of
Nauenberg (1972). On the other hand, interpolation in the tables of
Allen (1973) and allowing for the larger radius due to evolution
leads to a luminosity of $12.6\, L_\odot$ for the F-star in V1229~Cen. Since the
bolometric and the visual magnitude difference between the two system 
components will not be grossly different, the outbursting accretion disk 
should outshine the F-star by more than 4 magnitudes in visual light, 
in contrast to what is observed. 

This appears to be a serious problem if the outbursts are expected to be 
due to a thermal-viscous instability. Assuming the alternative, a temporarily
enhanced mass transfer from the donor star, it obviously vanishes. Moreover,
within the scenario of a hierarchical quadrupole system it is, of course,
also not existent.

\section{Conclusions}
\label{Conclusions}

Based on its optical light curve V1129~Cen has been classified as a 
$\beta$~Lyr type system. It distinguishes itself from other members of
this class by quasi-periodic eruptions, suggesting a relationship of the
star with dwarf novae. Here, I investigated its light curve in some detail
in order to either substantiate or reject this relationship.

Based on model calculations and comparisons with cataclysmic variables
it is concluded that the properties of V1129~Cen are not in contradiction
with the hypothesis that the system either contains, or that
it constitutes a dwarf nova, albeit with rather extreme characteristics.
In the first case V1129~Cen would be a hierarchical quadruple system, 
formed of two pairs, the optically dominating of which being a normal 
$\beta$~Lyr type variable. The second pair would be an ordinary dwarf nova.
However, while this scenario cannot {\it a priori} be discarded, such a 
configuration appears to be rather artificial.
Alternatively, V1129~Cen may consist of a Roche-lobe filling, slightly evolved
F2 star which loses mass via an accretion disk to a companion star; i.e.,
a cataclysmic variable with an unusually early type mass donor. The high
brightness of the F star is able to completely outshine the accretion disk
and the mass gainer (except during the occasional outbursts) such that the
normal photometric or spectroscopic hallmarks of CVs are not detected.  
A possible problem with this scenario arises if the outbursts are due to
a thermal-viscous instability (as opposed to a temporarily increased mass
transfer from the donor star) because then the accretion disk should become 
much brighter than observed.

As a caveat I stress that
this does not mean that the nature of V1129~Cen is elucidated beyond doubt.
I have shown that its properties are {\it compatible} with the hypothesis
that the system is a dwarf nova with a very early type mass donor, but it
would be premature to reject just for this reason another configuration 
for the star and alternative explanations for the outbursts.

\section*{Acknowledgements}

I gratefully acknowledge the use of the ASAS-3 data 
base which provided valuable supportive information for this study.

\section*{References}

\begin{description}
\parskip-0.5ex

\item      Allen, C.W. 1973, Astrophysical Quantities, third edition 
      (Athlone Press: London)
\item      Baptista, R. 2012, Mem.S.A.It., 83, 530
\item      Bateson, F.M. 1974, Publ.\ Var.\ Star Sect., RASNZ, 2, 1
\item      Brown, A.G.A, Vellenari, A., Prusti, T., et al. 2016, A\&A, 595, A2
\item      Bruch, A. 1993, 
      MIRA: A Reference Guide (Astron.\ Inst.\ Univ.\ M\"unster
\item      Bruch, A. 2016, New Astr., 46, 90
\item      Bruch, A. 2017a, New Astr., 52, 112
\item      Bruch, A. 2017b, New Astr., in press
\item      Bruch, A., \& Diaz, M.P. 2017, New Astr., 50, 109
\item      Caceci, M.S., \& Cacheris, W.P. 1994, Byte, May 1984, 340
\item      Catal\'an, S., Isern, J., Garc\'{\i}a-Berro, E., \& Ribas, I. 
      2008, MNRAS 387, 1693
\item      Claret, A., \& Bloemen, S. 2011, A\&A, 529, A75
\item      Dura\v{s}evi\'c, G., Latkovi\'c, O., Vince, I., \& Cs\'eki, A. 2010,
      MNRAS, 409, 329
\item      Dura\v{s}evi\'c, G., Vince, I., Antokhin, I.I., et al. 2012, MNRAS,
      420, 3081
\item      Eastman, J., Siverd, R., \& Gaudi, B.S. 2010, PASP, 122, 935
\item      Eggleton, P.P. 1983, ApJ, 268, 368
\item      Evans, P.A., Beardmore, A.P., Osborne, J.P., \& Wynn, G.A.
      MNRAS, 399, 1167
\item      FitzGerald, M.P. 1970, A\&A, 4, 234
\item      Garrido, H.E., Mennickent, R.E., Dura\v{s}evi\'c, G., et al. 2013,
      MNRAS, 428, 1594
\item      Gicger, A. 1987, Acta Astron., 37, 29
\item      Goliasch, J., \& Nelson, L. 2015, ApJ, 809, 80
\item      Harmanec, P., \& Scholz, G. 1993, A\&A, 279, 131
\item Hoffman, D.J, Harrison, T.E., Coughlin, J.L., et al. 2008, AJ, 136, 1067
\item      H{\o}g, E., Fabricius, C., Makarov, V.V., et al. 2000, A\&A, 355, L27
\item      Houk, N. 1978, Catalogue of two dimensional spectral types for the
      HD stars, Vol.\ 2, University of Michigan
\item Jacoby, G.H., Hunter, D.A., \& Christian, C.A. 1984, ApJ Suppl., 56, 257
\item      Kalomeni, B., Nelson, L., Rappaport, S., et al. 2016, ApJ, 833, 83
\item      Kazarovetz, E.V., Samus, N.N., Durlevich, O.V., Kireeva, N.N.,
      \& Pastukhova, E.N. 2008, IBVS 5863
\item      Kilkenny, D., \& Laing, J.D. 1990, SAAO Circ., 14, 11
\item      Kordopatis, G., Gilmore, G., \& Steinmetz, M. 2013, AJ, 146, 134
\item      Lasota, J.-P. 2001, New Astron.\ Rev., 45, 449  
\item      Mennickent, R.E., \& Djura\v{s}evi\'c, G. 2013, MNRAS, 432, 799
\item      Mennickent, R.E., Dura\v{s}evi\'c, G., Ko{\l}aczkowski, Z., \&
      Michalska, G., 2012, MNRAS, 421, 862
\item      Menzies, J.W., O'Donoghue, D., \& Warner, B. 1986, ApSS, 122, 73
\item      Nauenberg, M. 1972, ApJ, 175, 417
\item      Osaki, Y. 1996, PASP, 108, 39
\item      Pezzuto, S., Bianchini, A., \& Stagni, R. 1996, A\&A, 312, 865
\item Paczy\'nski, B., \& Schwarzenberg-Czerny, A. 1980, Asta Astron, 30, 127
\item      Pojmanski, G. 2002, Acta Astron, 52, 397 
\item      Rafert, J.B., \& Twigg, L.W. 1980, MNRAS, 193, 79
\item      Ritter, H. 2010, Mem.\ S.A.It., 81, 849
\item      Ritter, H., \& Kolb, U. 2003, A\&A, 404, 301
\item      Salaris, M., Serenelli, A., Weiss, A., \& Miller Bertolami, M.
      2009, ApJ, 692, 1013
\item      Shears, J., \& Poyner, G. 2009, JBAA, 120, 169
\item      Smak, J. 1983, ApJ, 272, 234
\item      Szkody, P., \& Mattei, J.A. 1984, PASP, 96, 988
\item      Walter, F., Bond, H.E., \& Pasten, A. 2006, IAU Cir.\ 8663
\item      Warner, B. 1987, MNRAS, 227, 23
\item Warner, B. 1995, Cataclysmic Variable Stars, Cambridge University Press,
      Cambridge
\item      Wilson, R.E. 1979, ApJ, 234, 1054
\item      Wilson, R.E., \& Devinney, E.J. 1971, ApJ, 166, 605
\item      Zacharias, N., Finch, C.T., Girard, T.M., et al.\ 2013, AJ, 145, 44
\item      Zhao, J.K., Oswalt, T.D., Willson, L.A., Wang, Q., \& Zhao, G. 2012,
      ApJ, 746, 144
\item Zorotovic, M., Schreiber, M.R., \& G\"ansicke, B.T. 2011, A\&A, 536, A42
\item      Zwitter, T., \& Munari, U. 1995, A\&AS, 114, 575
\item      Zwitter, T., \& Munari, U. 1996, A\&AS, 117, 449

\end{description}

\end{document}